\documentclass[a4paper, 10pt, twocolumn]{article}
\usepackage{amsmath, amsfonts, amssymb}
\usepackage{graphicx}
\usepackage{epstopdf}
\usepackage{epsfig}
\usepackage[ruled, lined, linesnumbered]{algorithm2e}
\usepackage{nomencl}
\usepackage{multicol}
\usepackage{color}
\usepackage{geometry}
\usepackage[toc, page]{appendix}

\newcommand{\dd}{\mathrm{d}}

\DeclareMathOperator{\prob}{\mathrm{Prob}}

\newcommand{\NC}[1]{{\textcolor{black}{#1}}}

\newcommand{\AStxt}[1]{\textcolor{black}{#1}}

\begin{document}

\title{Great cities look small}
\author{Aaron Sim \\ Department of Mathematics \& \\ Department of Life Sciences,\\ Imperial College London, SW7 2AZ, UK  \\ \\
\and Sophia N. Yaliraki \\ Department of Chemistry, \\Imperial College London, SW7 2AZ, UK \\ 
\and Mauricio Barahona\thanks{m.barahona@imperial.ac.uk}  \\ Department of Mathematics, \\Imperial College London, SW7 2AZ, UK  \\ \\
\and Michael P. H. Stumpf\thanks{m.stumpf@imperial.ac.uk}  \\ Department of Life Sciences, \\Imperial College London, SW7 2AZ, UK 
}
\twocolumn[
\begin{@twocolumnfalse}
\maketitle

\begin{abstract}
{Great cities connect people; failed cities isolate people. Despite the fundamental importance of physical, face-to-face social-ties in the functioning of cities, these connectivity networks are not explicitly observed in their entirety. Attempts at estimating them often rely on unrealistic over-simplifications such as the assumption of spatial homogeneity. Here we propose a mathematical model of human interactions in terms of a local strategy of maximising the number of beneficial connections attainable under the constraint of limited individual travelling-time budgets. By incorporating census and openly-available online multi-modal transport data, we are able to characterise the connectivity of geometrically and topologically complex cities. Beyond providing a candidate measure of greatness, this model allows one to quantify and assess the impact of transport developments, population growth, and other infrastructure and demographic changes on a city. Supported by validations of GDP and HIV infection rates across United States metropolitan areas, we illustrate the effect of changes in local and city-wide connectivities by considering the economic impact of two contemporary inter- and intra-city transport developments in the United Kingdom: {\em High Speed Rail 2} and {\em London Crossrail}. This derivation of the model suggests that the scaling of different urban indicators with population size has an explicitly mechanistic origin.\\ \\
}

\end{abstract}
\end{@twocolumnfalse}
]
{
  \renewcommand{\thefootnote}%
    {\fnsymbol{footnote}}
  \footnotetext[1]{m.barahona@imperial.ac.uk}
  \footnotetext[2]{m.stumpf@imperial.ac.uk}

}

\section{Introduction}
Can the greatness of a city be quantified? The city of Nineveh, capital of the Neo-Assyrian empire of 911-627 BC, was once described as ``an exceedingly great city, three days' journey in breadth" \cite{esv}. Today, a city described as such would more likely be dismissed as an urban sprawl let down by an inefficient transport infrastructure. Without reference to travelling-time constraints, size is clearly not a sufficient measure of greatness---just like rank and title can be poor predictors of influence in social networks \cite{watts, laidlawbook}. Of the many candidates \cite{marcb1, savitch}, the simplest objective measure of success is, possibly, the extent to which a city fulfils its primary purpose of maximising the number of face-to-face, opportunity-spawning, interactions between its inhabitants \cite{wu1}. From the rise of the Medici in ${15}^{\text{th}}$-century Florence to the prestige of an efficient transport system in a ${21}^{\text{st}}$-century metropolis, this \textit{connectivity} is synonymous with both the eminence of individuals and the success of whole cities \cite{batty2, padgettrobust, eagle1, granovetter, latora}.
\par
Measuring this connectivity, however, is not straightforward. Despite the success of social theory and experiments in much smaller contexts \cite{pentlandA, Basu, Cattuto}, the number of face-to-face social ties in a city, unlike secondary socio-economic indicators, remains poorly estimated. Beneath the reductionist representation of cities as featureless groups of individuals lies a forbidding, real-world diversity \cite{batty2}, including widely differing population sizes ($\sim$$10^3$--$10^7$), distributions (uniform, polycentric \cite{polycentric}), topologies and geometries, the latter covering both geography (boundaries, natural features) as well the different modalities of transport infrastructure (rail networks, traffic) \cite{Batty}. In addition, cultural and activity-specific behavioural differences (e.g. travelling-time tolerances) is a complicating factor in theories of urban human interactions.
\par
A typical strategy is to ignore this heterogeneity in favour of simple summary statistics like population size \cite{Bettencourt}, density \cite{Pan}, or even congestion sensitivity \cite{marcb1} or a global fractional dimensionality \cite{origins}. However, comparing cities that differ significantly on any of the excluded characteristics is then simply not possible with these models. Of particular significance to city planners, such models are, for the same reasons, unsuitable for assessing the impact of complex infrastructure or demographic changes.
\par
The parsimony of such approaches is, nevertheless, not without merit. Most notably, there is an apparent common scaling with respect to population size across a wide range of urban indicators \cite{Bettencourt2}. However, this empirical scaling is similar but not identical across indicators, both in the scaling exponent $\beta$ and level of statistical support (e.g. US 2002 new AIDS cases exhibits a power-law against population with an exponent $\beta = 1.23$ and correlation coefficient $\text{Adj-}R^2 = 0.76$  while private R\&D employment has $\beta = 1.34$ with $\text{Adj-}R^2 = 0.92$) \cite{Bettencourt}. Furthermore, power-law relationships can also arise by chance or as statistical artefacts, and even if supported by data they are largely descriptive and do not constitute constructive mechanistic narratives \cite{powerlaw, nee}. Indeed, recent attempts (such as in \cite{origins, arbesman, Toole:2015bl}) to lift this science of cities above the level of descriptive statistics reflect a growing desire for more generative and explanatory models. 
\par
A major step in this direction was taken by Pan et al. in \cite{Pan} where the observations behind the super-linear scaling relations were shown to be entirely consistent with -- and actually better modelled by -- the more fundamental assumption that the probability of social-tie formation between two individuals is inversely proportional to the number of people in closer proximity. In spite of the arbitrary nature of the probability ansatz, this elegant reduction of purely phenomenological power-law statistical observations to a statement about the likelihood of interactions between pairs of individuals suggests the existence of an underlying set of behavioural principles governing the formation of the network of social-ties in a city.
\par
In this paper we propose one such set of rules. These rules are `parameter-free' in the sense that they do not depend on any arbitrary functional assumptions beyond several intuitive statements on human behaviour. We build from them a model for real-world deliberate (as opposed to accidental or serendipitous) social interactions derived solely in terms of this set of agent-driven principles and is, therefore, by design, truly \emph{mechanistic}. In particular, via our derivation from first principles, we show how the probability of social-tie formation originally proposed in \cite{georouting} can be viewed as an emergent consequence of these more fundamental and, crucially, mechanistic principles. On a practical side, the model readily incorporates available detailed demographic, transportation and economic data, thereby providing a tool for the {\em a priori} assessment of the effectiveness of planned infrastructure measures. 

\section{A model of deliberate social ties}
\subsection{Modelling principles}
We start with four principles, the justification for and mathematical implications of which we will shortly unpack: 
\begin{enumerate}
\item
Individuals are characterized by a set of attributes \textit{(heterogeneity)}.
\item
For each attribute, individuals seek out social ties only with others who have higher attribute values \textit{(utility optimisation)}.
\item
Individuals have a set of attribute-specific travelling-time budgets $\tau_\text{max}$ \textit{(resource constraints)}.
\item
A directed tie is formed only if there are no closer and better opportunities in the proximity of the seeker \textit{(intervening opportunities)}.
\end{enumerate}
\paragraph{Heterogeneity}
The first principle is a nod to the variety of city life. Besides a multitude of attributes---from objective (e.g. wealth) to subjective (e.g. beauty), from beneficial (e.g. artistic skills) to harmful (e.g. criminality)---there exists a spectrum of skills and levels in those attributes across the population. \NC{To represent this heterogeneous set of attributes we define a set of non-identically distributed random variables}
\begin{equation}
\{X, Y, Z, \dotsc\}.
\end{equation}
Each set of realisations $\{x, y, z, \dotsc\}$ then represents an individual's set of abilities and scores in the corresponding attributes. 

\paragraph{Utility optimisation}
The second principle is a statement of human endeavour, whereby one seeks to build beneficial ties. It is simply a variation on the theory of rational choice where individuals are deemed to act in their own perceived best interest \cite{sugden}. For a given attribute $Z$, we express this necessary condition for a directed social tie from person $i$ to person $j$ as
\begin{equation}
(i \rightarrow j)_Z \Rightarrow z^{(j)} > z^{(i)}. \label{firstcond}
\end{equation}

\paragraph{Resource constraints}
The third principle reflects the finite nature of individual resources by adopting the concept of the travelling-time budget $\tau_\text{max}$, that is the maximum amount of time a person is willing to spend on a single commuting trip. There are several explanations for the key role it plays in the model. First, instead of Euclidean distances between geographical locations, a more faithful representation of a city's geometry is the set of real travelling-times along the spatially-embedded, multi-layered, transportation network between individuals (see, for example, \cite{Brockmann:2013bq}). Second, there is increasing evidence that the relevant measure for the formation of social ties is $\tau_\text{max}$ rather than the spatial separation between pairs of individuals (see~\cite{Mokhtarian} for a critical overview). In particular, it has been shown that in cities all across the world with high multimodal commuting behaviours, there is a uniformity in commute times that is independent of travel distance \cite{kung}.
\par
Here, instead of imposing a single, universal $\tau_\text{max}$, such as was done in \cite{Pan}, we allow for a list of different budgets $\tau_\text{max}^{X}, \tau_\text{max}^{Y}, \dotsc$ to reflect the heterogeneity of differing priorities and motivation levels for different activities undertaken by a single, fixed, population. For example, a city dweller who travels for three hours to attend an important business meeting might not be willing to spend more than 10 minutes on his weekly drive to a supermarket. 
\par
This principle gives us a necessary condition for the existence of a tie:
\begin{equation}
(i \rightarrow j)_Z \Rightarrow \tau_{ij} \leq \tau_\text{max}^Z, \label{secondcond}
\end{equation}
where $\tau_{ij}$ is the travelling-time distance between individuals $i$ and $j$.

\paragraph{Intervening opportunities}
The fourth principle represents the search heuristic that a person employs to perform constrained optimisation and is the defining geometric ingredient of our model. Each potential face-to-face interaction implies a minimal path defined by the shortest connecting travel route, which, in turn, defines a temporal social-sphere within which one evaluates the merit of the candidate interaction against other less costly options. These temporal spheres $S_{ij}$ are simply the sets of people that are closer to individual $i$ than another individual $j$, i.e. in a city of population size $N_\text{pop}$,
\begin{equation}
S_{ij} := \{k\, | \, \tau_{ik} < \tau_{ij}\}_{k=1}^{N_\text{pop}},
\end{equation}
with their cardinalities defining the components of the rank matrix\footnote{Note that, in general, $n_{ij} \neq n_{ji}$.}
\begin{equation}
n_{ij} := |S_{ij}|.\label{rankmat}
\end{equation}
Then, we can express a third necessary condition for a directed social tie as
\begin{equation}
(i \rightarrow j)_Z \Rightarrow z^{(j)} > \max_{k\in S_{ij}} z^{(k)}. \label{thirdcond}
\end{equation}
\par
In studies of human mobility, the consideration of such intervening opportunities has been shown to be the key to understanding travel patterns between cities \cite{Simini, intervening}. This fourth principle of our model is entirely consistent with and supports the growing body of evidence linking mobility and social contact patterns in cities \cite{Toole:2015bl}.
\par
As will be shown in the next section, these four principles, together with an assumption or prior knowledge of the spatial distribution of attribute values amongst the population, are sufficient to construct a weighted, directed network with the nodes $\{i, j, \dotsc\}$ representing a city's inhabitants and edge weights $\{\prob(i\rightarrow j)\}_{i,j = 1}^{N_\text{pop}}$ representing the probabilities of social ties between individuals. This probability network encapsulates the different levels of heterogeneity (attributes, geometry, topology, transport modality, and spatial population distribution) in our model of a city.  

From this probability network one can extract a host of statistics relevant to the problem at hand. Below we focus on the expected degree, i.e. the expected number of social ties of individuals in a city, which we take as a first measure of \emph{connectivity}, and which turns out to be a strong predictor for several urban indicators.

\subsection{Counting social ties}
By design of the model, the three conditions \eqref{firstcond}, \eqref{secondcond}, and \eqref{thirdcond} are together sufficient for the formation of the social tie $(i\rightarrow j)_Z$. The probability $\prob(i\rightarrow j)_Z$ is, therefore, simply the probability that those three conditions are satisfied.

We begin by setting $\tau_\text{max} \rightarrow \infty$, before reintroducing a finite $\tau_\text{max}$ at a later stage. Then by similar reasoning behind the radiation mobility model \cite{Simini}, we have 
\begin{equation}
\begin{split}
\prob(i\rightarrow j)_Z = &\prob(z^{(j)} > z^{(i)}) \,\times \\
&\prob\bigl(z^{(j)} > \max_{k\in S_{ij}} z^{(k)}\bigr).
\end{split}
\end{equation}
As we show in the \textit{Supp. Mat.} (see [S1]-[S5]), this equation can be simplified to give
\begin{equation}\label{dformula}
\prob(i\rightarrow j) = \frac{1}{n_{ij} + 2},
\end{equation} 
i.e. in the absence of travelling time budget constraints, the probability of a social tie is entirely determined by the rank matrix $n_{ij}$ \eqref{rankmat}, and is the same for all attributes (hence the dropped $Z$ label).

This probability expression is, for large $n_{ij}$, virtually equivalent to the proposal $\text{Prob}(i\rightarrow j) = 1/{n_{ij}}$ as introduced in \cite{georouting} and developed in \cite{Pan}. Crucially however, we have shown that it can in fact be derived directly from first principles, and is naturally regularised being well-defined when $n_{ij}=0$ without the need for artificial and arbitrarily imposed constraints on the minimum sizes of social-spheres (see \cite{Pan}). Remarkably also, the attribute-dependency retained at the beginning of our derivation drops out naturally from the final expression -- our model is, therefore, a non-trivial instance of a probabilistic and mechanistic social interaction model consistent with observations of emergent urban-feature independence \cite{Bettencourt}. 
\par
Clearly, the key input of the model is, then, the travelling-time distance matrix $\tau_{ij}$ from which one uses to build the rank matrix $n_{ij}$. The data required for constructing $\tau_{ij}$ are often public and readily available online through a variety of tools\footnote{e.g. Google Distance Matrix API, MapQuest Route Matrix, Microsoft Bing Routes API.}, as demonstrated in the application examples in Section \ref{applications}. 
\par
The expected total number of ties $T_Z$ corresponding to an attribute $Z$ in a population of size $N_\text{pop}$ is then simply the sum over each individual set of probabilities up to a finite $\tau_\text{max}^{Z}$, i.e.
\begin{equation}
T_Z = \sum_{i,j=1}^{N_\text{pop}} \frac{1}{n_{ij} + 2} \mathbb{I}(\tau_{ij} \leq \tau_\text{max}^Z).
\end{equation}

Although technically correct, building the distance matrix $\tau_{ij}$ covering the entire population is highly impractical for all but the smallest of cities. Instead, we subsample the geographical extent of the city at  $N_\text{s}\, (\ll N_\text{pop})$ points to generate the much smaller sample distance matrix ${\hat{\tau}}_{ij}$. From this coarse-grained representation of the city, we obtain the approximation
\begin{equation}
\label{approxform}
T_Z \approx N_\text{pop}\left[\ln \left(\frac{N_{\text{pop}}}{2N_\text{s}}\right) + \frac{1}{N_\text{s}}\sum_{i=1}^{N_\text{s}}\ln {n}_i ^Z\right]+ \frac{2N_s}{\bar{n}^Z},
\end{equation}
where ${n}_i^Z := \sum_{k=1}^{N_\text{s}} \mathbb{I}({\hat{\tau}}_{ik} \leq \tau^Z_\text{max})$ is the size of the social sphere, as related to attribute $Z$, of the location $i$ in the subsampled city, and $\bar{n}^Z = (1/N_s)\sum_{i=1}^{N_s} n_i^Z$ (see \textit{Supp. Mat.} for the derivation of this approximation). In the following section, we show through a series of simulations that this approximation is both unbiased and robust.  

For the remainder of the paper, we drop the $Z$ label for notational clarity.

\subsection{Local connectivity}
The total number of ties $T$ is a global, city-wide, connectivity measure which encapsulates the intricate complexities of the city geometry and heterogeneities in agent attributes. Our model also offers a measure that captures the spatial variation in tie-formation across a city. We introduce the concept of the local connectivity of some sub-region of a city as the sum of all incoming and outgoing ties. Let $T_i$ represent the local connectivity at the location of individual $i$, such that $T=\sum_{i=1}^{N_\text{pop}} T_i$. Then
\begin{equation}\label{localform}
\begin{split}
T_i &= \frac{1}{2}(T_i^\text{from} + T_i^\text{to})\\
&=\frac{1}{2}\ln \left(\frac{\alpha n_i}{2} +1\right) + \frac{\gamma\alpha}{2}\sum_{\substack{j=1\\j\neq i}}^{N_s}\frac{1}{\alpha\bigl(\hat{n}_{ji} + \frac{3}{2}\bigr) + \frac{1}{2}},
\end{split}
\end{equation}
where $\alpha = N_\text{pop}/N_s$ and $\gamma$ a scaling factor that ensures, for consistency, that $\sum_{i=1}^{N_s} T_i^\text{from} = \sum_{i=1}^{N_s} T_i^\text{to}$ (for a full derivation see {\em Supp. Mat.}).
\par
The distribution of $T_i$ reflects the heterogeneity of the induced interaction network (see \textit{Supp. Mat.} Figure S3d). In particular it enables one to quantify the distinct and disproportionate influence that transportation and other infrastructure schemes can have in different parts of the city, as we show in an example in Section \ref{crossrail} below.

\subsection{Relating social-tie connectivity with other measurable indicators}
Our underlying assumption is that there is a link between the attribute-specific social-tie connectivity $T$, as defined in \eqref{approxform}, and a measure $U$ of a related productive urban activity: 
\begin{equation}
\label{eq:U_full}
U = f(T) = a_0 + a_1T + a_2T^2 + \dotsb.
\end{equation}
$U$ can correspond to socio-economic measures such as GDP, innovation indices, etc. We are primarily interested here in scenarios where the contribution of individual, isolated, efforts is either non-existent (e.g. spreading of disease) or negligibly small (e.g. collaborative scientific research output). In such cases, $a_0 =0$. As a first approximation, we consider here a simple proportional relation with $a_{i>1} = 0$, which often provides reasonably good explicative power (see \cite{Pan} and \cite{burchardi}). For example, if the probability $p$ of disease transmission in a single encounter between an infected and susceptible individual is small (e.g. sexual per-act HIV transmission risk is $<0.014$ \cite{Patel:2014br}), then within a relatively short timeframe the total number of new infection cases given $T$ such interactions is simply $pT$. We, therefore, define our relation to be simply
\begin{equation}\label{relation}
U = aT,
\end{equation}
with $a\in \mathbb{R}$ the single unknown parameter relating connectivity and its related activity measure. In situations where the first-order approximation breaks down, the networks of social ties generated through our model allow the use of higher statistics beyond the average degree, which could be used to test hypotheses against~\eqref{eq:U_full}. We discuss this point further at the end of the paper (see also \textit{Supp. Mat.} where we discuss the expected degree distribution).

In summary, there are just two parameters in the model: the constant of proportionality $a$ and, implicit in the computation of $T$, the travelling time budget $\tau_\text{max}$. We emphasise that these parameters have precise meanings in the model, i.e., they are not just \textit{post hoc} adjustable tuning levers,  and that they can be inferred from data to characterise the dynamics and the implications of human interactions contained in the observations (for an example, see Section \ref{evidenceattdep}). Alternatively, the parameters, $\tau_\text{max}$ in particular, can be fixed using prior knowledge, such as from travel behaviour surveys, information from similar cities, or from crowd-sourced location-data. Furthermore, under the linear assumption, the typical exercise of comparing scenarios (e.g. the relative increase of economic activity before and after the completion of a new railway) affords a further simplification, as the parameter $a$ cancels out when taking ratios. 

\section{Validation of the social-tie model}
The mathematical model above formalises the hypothesis-driven narrative stemming from
our set of  agent-driven, behavioural principles, and represents a possible mechanistic process of face-to-face communication within a general population together with its city-level phenomenological implications. To check the implications of the model, we have performed a set of simulations and empirical validations. 

We begin by validating the procedure to obtain $T$, the total number of ties. There are two separate aspects to consider: \textit{(i)} the statistical validity of the sampling approximation \eqref{approxform} for the population-level $T$; and \textit{(ii)} the validity of the rank-based formula \eqref{dformula} for the probability of a tie between two individuals given the four principles in our model. We examine both parts together in a single set of simulations, as described below.

\subsection{Statistical surrogates of cities with multi-modality mobility}
To test our model, we generate multiple surrogates of cities and the corresponding travelling-time matrices under multi-modal transport networks. These simulated cities are designed to model real-world urban mobility patterns involving multiple transport modes. We consider four population sizes $N_\text{pop}=(300, 500, 800, 1200)$, with five different population distributions (a uniform distribution over a $45 \times 45 $ km square area, and a 2-dimensional, circularly symmetric, Gaussian distribution with standard deviations of 3, 6, 9, and $12$ km) and two travelling time budgets ($\tau_{\text{max}} = 1, 2$ hours). 

\begin{figure*}
\begin{center}
\centerline{\includegraphics[width=15cm]{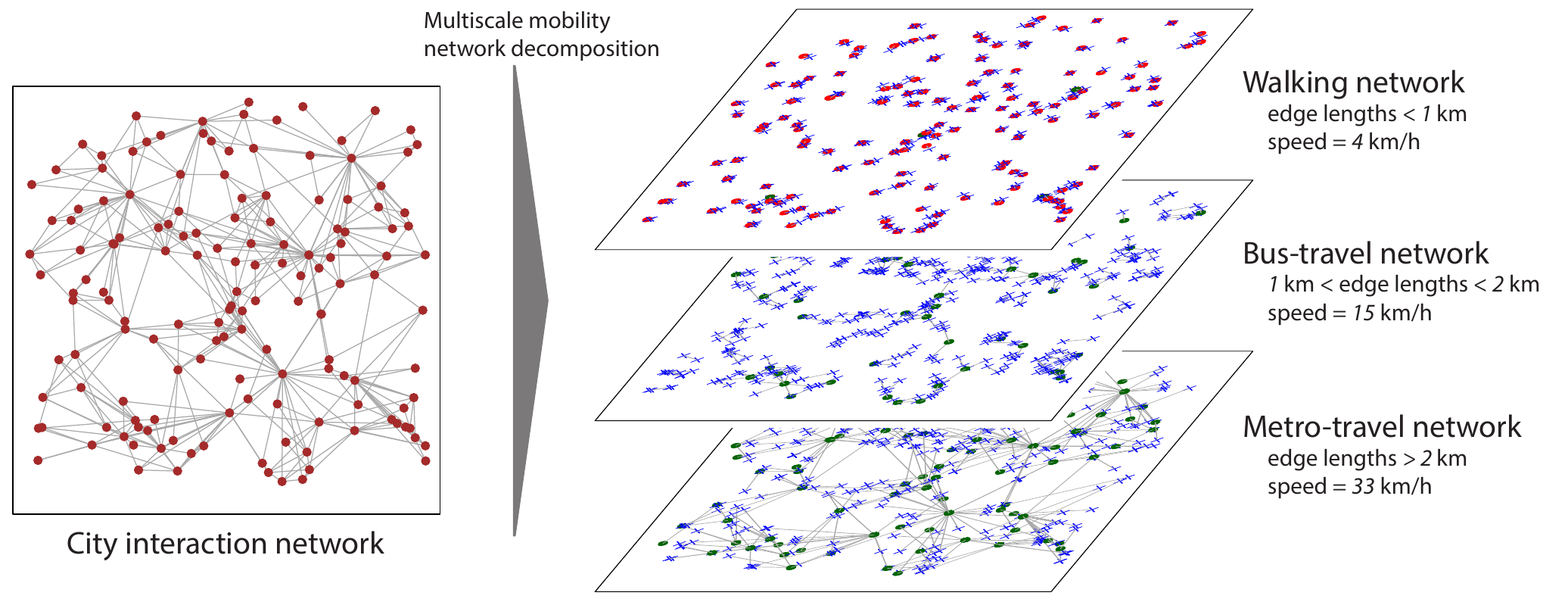}}
\caption[Network decomposition]{\textbf{Multilevel mobility network decomposition of urban interaction networks}. In the multilayer mobility networks, the red and green nodes represent the origin and destination, respectively, of the particular directed edge in the city interaction network. The blue crosses indicate a transfer from one transport mode to another (e.g. walking to metro), where each cross on a given layer corresponds to another on a different layer. Note that the spatial position of each transfer node in each layer have no meaning other than to provide an indication of the spatial distance travelled in the corresponding mode. }
\label{networkdecomp}
\end{center}
\end{figure*}

To simulate the multi-modal transportation infrastructure we proceed as follows. For each pair of
individuals $i, j$ in our simulated city, we compute the Euclidean spatial distance $s_{ij}$ and decompose into binary form:
\begin{equation}
s_{ij} \equiv \bigl(s_{ij}^{(0)} \cdot 2^0\bigr) + \bigl(s_{ij}^{(1)} \cdot 2^1\bigr) + \bigl(s_{ij}^{(2)} \cdot 2^2\bigr) + \dotsb,
\end{equation}
where $s_{ij}^{(k)} \in \{0,1\}$. The multi-modality transport network is represented by a speed vector $\mathbf{v} = (v_0, v_1, \dotsc, v_m)$, where each component  is the speed of a certain transportation mode in order of increasing speed,  $v_{k+1} \geq v_k$. We then generate the travelling-time distance matrix $\tau_{ij}$ between all pairs of points in the city as
\begin{equation}\label{simdist}
\tau_{ij} = \sum_{k=1}^m \frac{s_{ij}^{(k)}\cdot 2^k}{v_k}.
\end{equation}
This framework for the simulation of travelling-times replicates two features of modern-day transport infrastructure, which is illustrated in Figure \ref{networkdecomp}. First, there is the hierarchical nature of travelling speeds with faster transport modes covering larger distances. Second, the framework allows for the fact that travel between two locations in a city typically involves a combination of transport modes (e.g. bus + train). The slowest mode of transportation is given by $v_0 = 4\,\mathrm{km/h}$. A city with no transport infrastructure will be represented by a vector $\mathbf{v} = (4,\dotsc, 4)$ and the time between nodes is then the time taken to walk the spatial separation distance. A more realistic case, where public transportation modes of walking, bus and train networks are considered is represented by $\mathbf{v} = (4, 10, \dotsc, 100)$).  If private travel is considered, different classes of roads and expressways traversed using bicycles or automobiles could be considered. In our simulations, we considered four different transport infrastructures, as shown in Table \ref{travelspeeds}.

In summary, four population sizes, five distributions, two travelling time budgets, and three non-trivial transportation infrastructures give a total of 120 unique surrogate cities, each given by its specified distribution of $N_\text{pop}$ points on a square $45 \times 45\, \mathrm{km}$ grid and a resulting $N_\text{pop} \times N_\text{pop}$ travelling-time distance matrix $\tau_{ij}$. 

\subsection{Validation of the sampling procedure and probability model}
To validate our sampling~\eqref{approxform}, we compare the travelling-time distance matrix~\eqref{simdist} in our simulated cities obtained from the whole population $N_\text{pop}$ and from a reduced sample of $N_s=150$ points, as follows. Every one of the $150\times 149=22350$ possible directed ties in the sample is assigned a probability according to \eqref{dformula}. The total number of ties in the sample is obtained by summing over the probabilities, which are then scaled up according to \eqref{approxform}. 

In the simulation of the full population $N_{\text{pop}}$, we take the viewpoint of each individual, and we rank the other $N_\text{pop} -1$ people in the population according to their travelling-time distances from the individual. We consider a population characterized by an attribute, and the individuals are i.i.d. instances drawn from a standard log-normal distribution. There are $N_\text{pop}(N_\text{pop}-1)$ possible directed ties. Starting from the closest person, a directed tie from the individual is assigned according to the fourth modelling principle of intervening opportunities subject to the upper constraint of an upper bound $\tau_{\text{max}}$ for the travelling time.

The results of the comparison between the full population and the sample are shown in Figure \ref{US}a and the close match demonstrates the validity of the probability model \eqref{dformula} as well as demonstrating that the sampling procedure \eqref{approxform} provides a good and unbiased approximation.

\subsection{Comparison with power-law scaling models}
\begin{figure*}
\begin{center}
\centerline{\includegraphics[width=15cm]{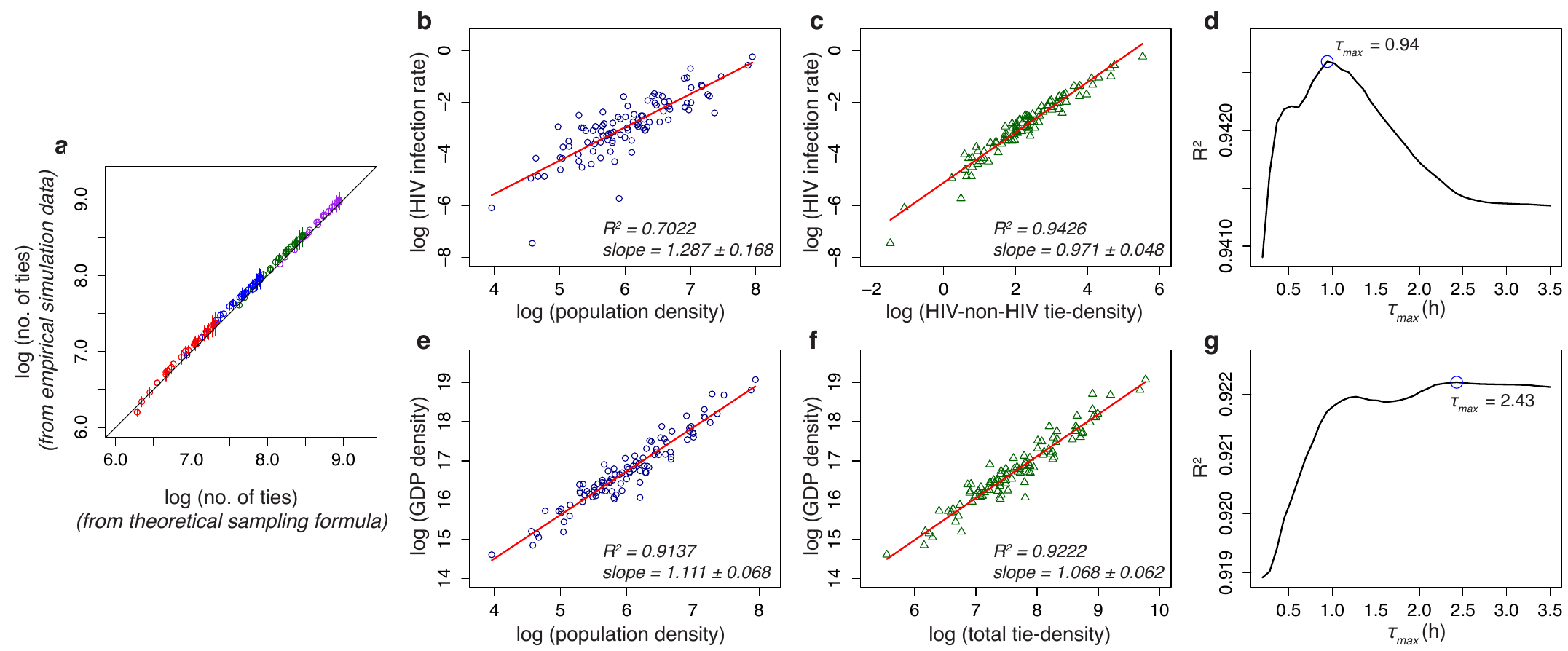}}
\caption[Validation on US data]{\textbf{Validation of sampling procedure and empirical validation with HIV infection rates and GDP of 102 US Metropolitan Statistical Areas}. \textbf{a}, Comparison of the total number of ties empirically counted according to the interaction model \textbf{(y-axis)}, with the number of ties estimated from population samples of 120 simulated cities, according to \eqref{approxform} and \eqref{dformula} \textbf{(x-axis)}. The four colours (red, blue, green purple) indicate population sizes of 300, 500, 800, and 1200 respectively. Further variation in the cities are created by imposing different population distributions, maximum travelling-time budgets, and transport infrastructure. The circles indicate the mean of 30 simulations and the vertical lines $\pm 2$ standard deviations. As shown, the sampling procedure provides a reasonably good estimate of the total number of ties. \textbf{b, e}, Power-law fits of urban indicators to population density. \textbf{c, f}, Linear fits of urban indicators to tie-density with $\tau_\text{max}$ set at the maximum likelihood values (as indicated by the blue circles in \textbf{d, g}). \textbf{e, g}, Coefficient of determination of tie-density fits as a function of maximum travelling-time budget $\tau_\text{max}$. The error values on the slope parameters indicate $\pm 2$ standard deviations. We note that for both urban indicators, the fits to total tie-density outperforms the fits to population density.}\label{US}
\end{center}
\end{figure*}

Using real-world data from US cities, we compare the predictive abilities of our model and that of power-law scaling models~\cite{Bettencourt}. We begin by generating travelling-time distance matrices on sampled representations of 102 United States Metropolitan Statistical Areas (US MSAs). The detailed information available\footnote{\textit{2010 Census of Population and Housing} \& \textit{2010 U.S. Metropolitan Statistical Area Distance Profiles}, www.census.gov; www.microsoft.com/maps/} on the population distributions in these MSAs allows us to construct sample distance matrices that are representative of the full population-scale distance matrices. We then plot the computed number of social ties $T$ (as a function of the travelling-time budget $\tau_\text{max}$) from our model against two measures of urban activity $U$: the 2011 gross domestic product (GDP) and HIV infection rate\footnote{US Centers for Disease Control and Prevention. \textit{HIV Surveillence Report}, 2011; vol.23. www.cdc.gov/hiv/topics/surveillance/resources/reports/. Feb 2013.}. We also make the comparison with the corresponding power-laws against population density. As shown in Fig. \ref{US}, the model is, on its own, well supported by the data with a linear $\log U$-$\log T$ relationship with $\text{slope}\approx 1$. Our social-tie model provides an equally good fit for the GDP case ($R^2 = 0.92$ (social-ties) vs. $0.91$ (power-law)) and has a significantly stronger statistical support compared to the power-law fit to population density in the HIV infection rate case ($R^2 = 0.94 \text{ vs. } 0.70$). Much of this improvement stems from the shift from counting people to counting ties -- specifically ties between HIV-positive and negative individuals (see \textit{Supp. Mat.}). It is the overly-broad category of a city's economic output and the lack of specificity in the nature of such relationships that explains the relatively marginal improvement in statistical support in the GDP example. Together, the examples support the view that the fundamental units of a city are not its inhabitants but the social relationships that exist between them. 

\subsection{Evidence for the attribute-dependence of the travelling-time budget}\label{evidenceattdep}
In addition to its predictive performance shown above, and because of its agent-driven construction, our model can also shed light on the mechanistic origin of social interactions. For instance, the two examples above (GDP and HIV infection) highlight a marked difference in the underlying social dynamics across the two attributes considered, as seen from the corresponding maximum likelihood estimates of $\tau_\text{max}$. We obtain $\tau_\text{max}=2.43\,\mathrm{h}$ (95\% C.I. $[0.36\,\mathrm{h}, 5.42\,\mathrm{h}]$) for the GDP output versus a markedly lower value of $\tau_\text{max}=0.94\,\mathrm{h}$ (95\% C.I. $[0.36\,\mathrm{h}, 1.52\,\mathrm{h}]$) for HIV infection rates. The confidence intervals are given by quantiles from bootstrapped samples of the original data set (see \textit{Supp. Mat.}).

Ignoring for the moment the small range of variation in $R^2$ values with $\tau_\text{max}$, there are two immediate interpretations. First, our fits indicate that, in contrast to economically productive activities, it is unlikely that one would be willing to travel for more than 1.5 hours to engage in activities associated with HIV transmission. Second, as expected, GDP stems from a wide range of activities leading to a more variable $\tau_\text{max}$. Recognising and quantifying such differences in interpretable parameters and their variances, which would be missed by simple scaling arguments, is of relevance in efforts to build both prosperous and healthy cities.

Nevertheless, despite the bootstrapped analysis giving confidence intervals for our $\tau_\text{max}$ estimates, the small range of variation in $R^2$ suggests a level of redundancy in our model with the constant of proportionality $a$ in \eqref{relation} affording too much freedom. In order to increase the robustness of the model when applied to real data, we eliminate the proportionality parameter $a$ by considering relative increases of indicators, i.e., we consider the ratio $U_1/U_2$ of the economic indicators. This is illustrated in the next section, where we provide two examples of the application of this approach.

\section{Applications of the social-tie model}\label{applications}
To illustrate the applicability of our model, we examine two examples of large-scale transportation projects in the United Kingdom: High Speed Rail 2 (HS2) and London Crossrail. 

\subsection{The High Speed Rail 2 project}
\begin{figure*}
\begin{center}
\centerline{\includegraphics[width=15.0cm]{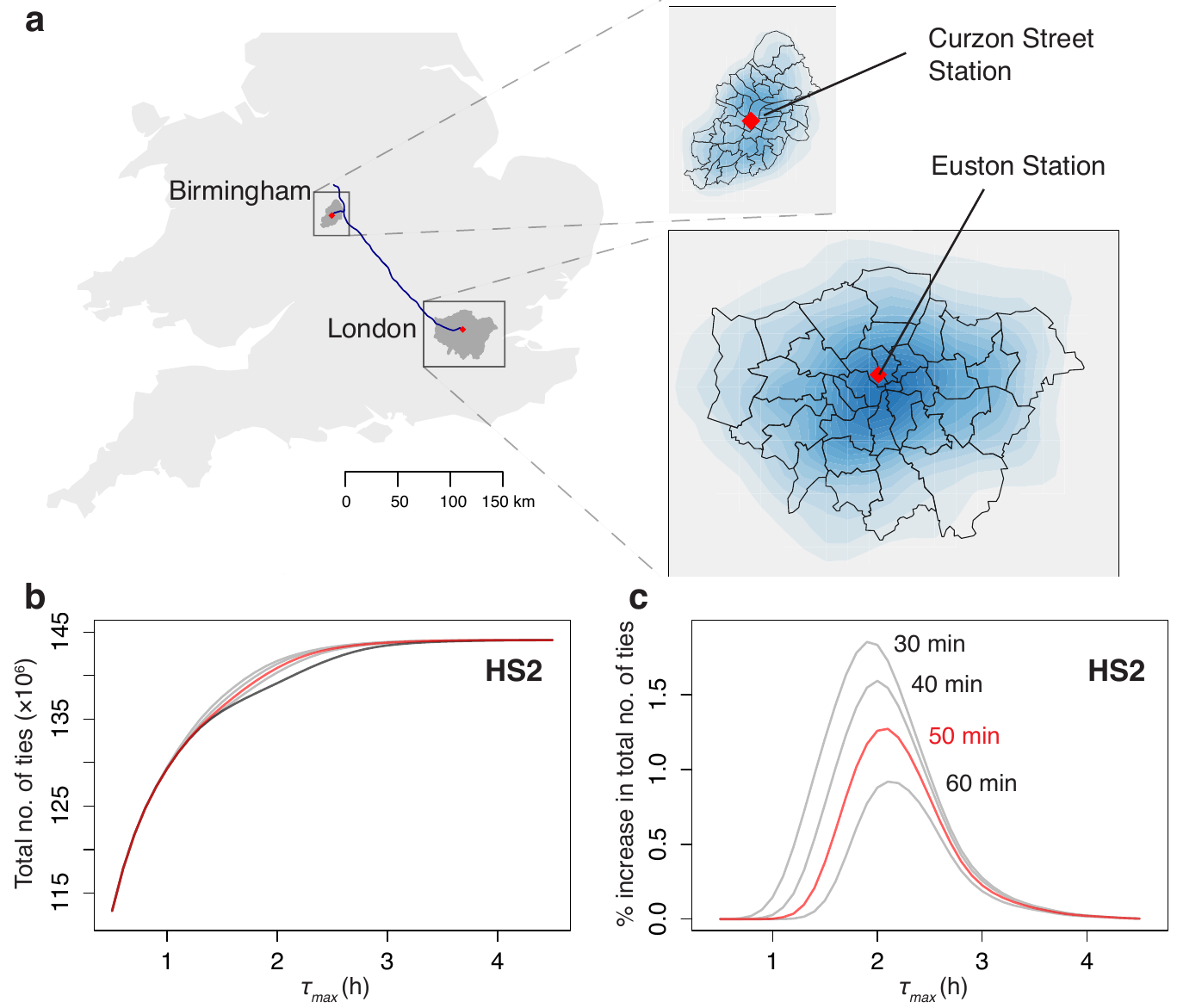}}
\caption[High Speed Rail 2]{\textbf{High Speed Rail 2 (Phase 1) and its impact on the connectivity of UK cities}. \textbf{a,} High Speed Rail (Phase 1) route and the population densities of London and Birmingham. The blue line indicates the published proposed route of the first phase of HS2 (as of Dec-2013). The red diamonds indicate the locations of the rail stations in each city. The contour maps are derived from kernel density estimates of 1000 and 129 samples points in London and Birmingham respectively. The ratio of the number of samples is chosen to reflect the relative sizes of the two cities. \textbf{b,c,} Impact of High Speed Rail 2 (Phase 1) on the connectivity of UK cities. The black curve indicates the connectivity without HS2. The red curves indicate the connectivity according to the planned improved travel times (50 mins between London and Birmingham). The grey curves in \textbf{c} indicate hypothetical travel times of 30, 40 and 60 minutes.}\label{HS2map}
\end{center}
\end{figure*}

HS2 is the proposed high-speed rail network connecting the major cities in Britain, from London in the south to the northern cities of Leeds, Manchester and beyond. In this section we focus on the first phase link between London and Birmingham that would reduce the one-way travel time from the current 84 to 50 minutes. We treat the two cities as a single conurbation and omit the influence of the neighbouring regions; the results presented here should be interpreted in the light of this geographical treatment. In Figure~\ref{HS2map} we plot the total and percentage increases in the number of ties as a function of $\tau_\text{max}$. If we take the value of $\tau_\text{max}=2.43\,\mathrm{h}$, which we inferred previously for the GDP-related travelling-time budget, the average economic boost induced by the presence of HS2 across the two cities would be $\approx 0.96\%$. A more robust approach is to consider a range of possible time-budgets to evaluate the effect of uncertainty in $\tau_{\text{max}}$ (see \textit{Supp. Mat.}). For instance, assuming a uniform distribution over $1 < \tau_\text{max} < 3$ we obtain an increase in GDP of 0.80\%. Interestingly, we observe a middle `sweet spot' at $\tau_\text{max}\sim 2\,\mathrm{h}$: at the lower tail, the journey times are insufficiently short to tempt one to travel further, while at the upper tail, the efforts are wasted on a population already willing to endure long commutes. 

\subsection{London Crossrail}\label{crossrail}
\begin{figure*}
\begin{center}
\centerline{\includegraphics[width=15.0cm]{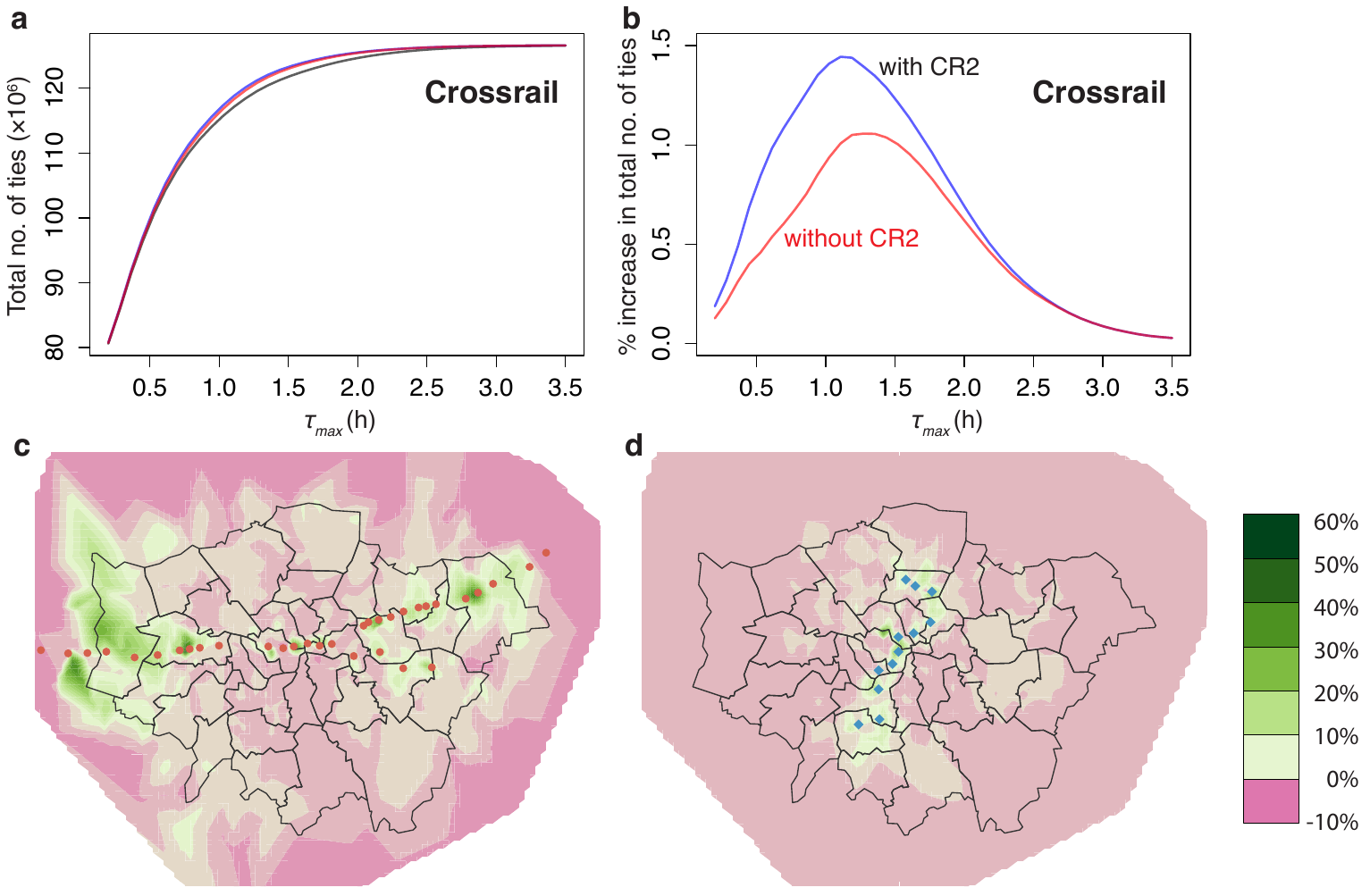}}
\caption[Crossrail and local connectivity]{\textbf{Impact of London Crossrail on city-wide and local connectivities}. \textbf{a,b,} Impact of London Crossrail on the connectivity of London. The black curve indicate the present connectivity without Crossrail. The red curves indicate the connectivities according to the planned improved travel times from Crossrail (but without Crossrail 2). The blue curve in \textbf{b} shows the connectivity boost by including Crossrail 2 (metro-only option), a proposed project extension to include a North-South train link. \textbf{c}, Percentage change in local connectivity due to Crossrail. \textbf{d}, Percentage change in local connectivity due to London Crossrail 2 (metro-only option) relative to post-Crossrail. The heat map scales indicate percentage change in the total number of incoming and outgoing ties for each region. The red points indicate the Crossrail stations and the blue points the 12 proposed Crossrail 2 stations.}\label{CRlocal}
\end{center}
\end{figure*}

\emph{Crossrail} is a high-frequency railway linking east and west London currently under construction. Under the same $\tau_\text{max}$ assumptions as for HS2 above, the projected impact of Crossrail on the London economy is a $0.3\%$ increase in the city's GDP (with an increase of 0.61\% for the uniform distribution of  $\tau_\text{max}$) (Fig. \ref{CRlocal}). The percentage increases may appear small ($<1\%$), but it is by no means unexpected for two reasons. First, the stated investment cost is itself a small fraction of London's GDP. Second, the modest boost is simply a reflection of the highly concentrated population density in the central regions and the extensive transport infrastructure already in place.

The availability of precise local geographical data allows us to further interrogate the model to determine the spatial distribution of local connectivities $T_i$ \eqref{localform}. Indeed, it is important to note that neither the current local connectivity levels nor the impact of Crossrail are evenly distributed or felt across the city (see Fig. \ref{CRlocal}). As would be expected, the largest increases are found near railway stations, especially in London's suburbs. As we explore further in \emph{Supp. Mat.} (see Fig. S5), there is a concentration of newly possible connections along the east-west extent of the city. More surprisingly, however, we observe a \textit{decrease} across large areas along the orthogonal north-south axis driven by falls in their \textit{relative} accessibility---the rising tide of connectivity does not lift all boats. This effect may be unavoidable, but the ability to quantify and map its spatial extent allows one to anticipate and, possibly, alleviate its impact. 

There is a mooted north-south extension -- \emph{Crossrail 2} --  which is currently under study (see \textit{Supp. Mat.} for details). In similar fashion to Crossrail, the expected additional boost to GDP can be calculated and is shown in Figure \ref{CRlocal}. Crucially, in line with one's intuition, the negative local impact is now distributed outside the areas surrounding the Crossrail 2 rail line. 

\section{Discussion}
Unlike typical social network and epidemiological studies that assume a fixed and known network structure within which various dynamical processes (e.g. spread of diseases) are constrained, our approach obtains interaction networks as induced structures that emerge from the application of our set of principles to different cities. In this sense, these interaction networks are unobserved structures, much like genealogical trees in population genetics \cite{Rosenberg:2002ff}. Unlike random geometric graphs emerging in models of cities with uniform population distributions~\cite{mattpenrose}, our model incorporates agent-driven optimisation principles and physical constraints from the geometry and topology of each city. Hence, rather than functioning as input features for our model, these resulting networks capture and are confined by the make-up of the demographic and transport infrastructure data under study. 
\par
Although the unobservable nature of the underlying connectivity networks poses challenges for the direct validation of our model, the recent availability of large-scale location data from mobile phones appears to offer a wealth of possibilities for testing some of the model assumptions, e.g., the existence of travelling-time budgets $\tau_\text{max}^Z$, and their assumed uniformity across the population for each attribute. However, there are specific conditions that such empirical studies must fulfil. In particular, one should be able to identify, with reasonable certainty, the purpose and deliberateness of both single journeys and social ties observed.  In this context, the growth of location-based and, crucially, activity-specific, social networking services could provide valuable information \cite{Cho:2011io}, in contrast to simply relying on proximity information for social tie prediction \cite{Wang:2011fz}.
\par
As shown above, the overall connectivity $T$ is, on its own, a strong predictor for several urban indicators and we have concentrated on this aspect in this paper. This is reassuring given the known ability of mean-field theory to capture basic trends~\cite{gleeson} on networks. Nevertheless, further details and statistics (e.g. heterogeneity) of the obtained networks could be studied, as the mechanistic and constructive nature of our model provides the necessary information for extracting these additional features. We provide a short illustration of this process in the \textit{Supp. Mat.}. An extension of our model will be to propose and test the analogue of \eqref{relation} with different network statistical measures in place of $T$. 
\par
The generic nature of the proposed framework and the increasing availability of geo-location and travel data ensure a broad and growing array of applications. This includes gauging the robustness of a city to traffic congestions and measuring the cost of weather-related disruptions. Methodological extensions to the model might include, for instance, replacing travel time with a cost function incorporating spatial distance, financial cost and the time-of-day. 
\par
\AStxt{Our focus for most of this paper has been on the city as defined by civil administrative conventions. Since studies of cities are sensitive to the exact definition of a city itself \cite{Rozenfeld:2009je, Oliveira:2014gz}, there is the option of adopting one of the more nuanced alternative definitions that do not include any arbitrary geographical boundaries \cite{Rozenfeld:2008et}.  However, the model itself is actually agnostic as to the source of the population variables $N_\text{pop}$ or the travelling-time distance matrices $\tau_{ij}$, as indeed we have shown by treating the two cities of London and Birmingham as a single entity in our analysis above. Our approach can thus be applied to reflect the connectivity among geographic entities both on a larger scale (countries or larger geographical regions) and a smaller scale  (buildings or campuses). On such smaller scales, this approach can inform design to maximise the creative, social and economic benefits resulting from human encounters. Regardless of the context of application, it is not the actual spatial size but the extent \textit{perceived} via travelling times that determines the connectivity of a system. Large cities may be great, but great cities most certainly look small.}

\section*{Author Contributions}
AS, SY and MB conceived the study; AS produced the theoretical material, collected and analysed the data, and drafted the manuscript. MPHS coordinated the study. All authors participated in the design of the study, helped draft the manuscript and gave final approval for publication.

\section*{Funding statement}
This project has been funded by EPSRC grant number EP/I017267/1 under the \textit{Mathematics Underpinning the Digital Economy} Programme.

\bibliography{STies_26.bib}

\begin{thebibliography}{10}

\bibitem{esv}
{\em {The English Standard Version Bible, Jonah 3:3}}.
\newblock Crossway, 2001.

\bibitem{watts}
D~J Watts and P~Dodds.
\newblock {The accidental influentials - Microsoft Research}.
\newblock {\em Harvard Business Review}, 85(2):22--23, 2007.

\bibitem{laidlawbook}
Zo{\"e} Laidlaw.
\newblock {\em {Colonial connections, 1815-45 : patronage, the information
  revolution and colonial government}}.
\newblock Manchester, UK ; New York : Manchester University Press ; New York :
  Distributed exclusively in the USA by Palgrave, 2005.

\bibitem{marcb1}
R{\'e}mi Louf and Marc Barth{\'e}lemy.
\newblock {From mobility patterns to scaling in cities}.
\newblock {\em arXiv.org}, January 2014.

\bibitem{savitch}
H~V Savitch.
\newblock {What makes a great city great? An American perspective}.
\newblock {\em Cities}, 27(1):42--49, January 2010.

\bibitem{wu1}
Lynn Wu, Ben Waber, Sinan Aral, Erik Brynjolfsson, and Alex Pentland.
\newblock {Mining Face-to-Face Interaction Networks Using Sociometric Badges:
  Predicting Productivity in an IT Configuration Task}.
\newblock {\em International Conference on Information Systems}, 2008.

\bibitem{batty2}
Michael Batty.
\newblock {\em {The New Science of Cities}}.
\newblock MIT Press, November 2013.

\bibitem{padgettrobust}
John~F Padgett and Christopher~K Ansell.
\newblock {Robust Action and the Rise of the Medici, 1400-1434}.
\newblock {\em American Journal of Sociology}, 98(6):1259--1319, May 1993.

\bibitem{eagle1}
Nathan Eagle, Michael Macy, and Rob Claxton.
\newblock {Network diversity and economic development}.
\newblock {\em Science (New York, N.Y.)}, 328(5981):1029--1031, 2010.

\bibitem{granovetter}
Mark Granovetter.
\newblock {The Impact of Social Structure on Economic Outcomes}.
\newblock {\em Journal of Economic Perspectives}, 19(1):33--50, January 2005.

\bibitem{latora}
Vito Latora and Massimo Marchiori.
\newblock {Efficient Behavior of Small-World Networks}.
\newblock {\em Physical review letters}, 87(19):198701, October 2001.

\bibitem{pentlandA}
Alex~Sandy Pentland.
\newblock {Automatic mapping and modeling of human networks}.
\newblock {\em Physica A: Statistical Mechanics and its Applications},
  378(1):59--67, 2007.

\bibitem{Basu}
S~Basu, T~Choudhury, and B~Clarkson.
\newblock {Towards measuring human interactions in conversational settings}.
\newblock {\em Proc IEEE CVPR}, 2001.

\bibitem{Cattuto}
Ciro Cattuto, Wouter Van~den Broeck, Alain Barrat, Vittoria Colizza,
  Jean-Fran{\c c}ois Pinton, and Alessandro Vespignani.
\newblock {Dynamics of person-to-person interactions from distributed RFID
  sensor networks.}
\newblock {\em PloS one}, 5(7):e11596--e11596, January 2010.

\bibitem{polycentric}
R~Louf and M~Barthelemy.
\newblock {Modeling the Polycentric Transition of Cities}.
\newblock {\em Physical review letters}, 111:198702, November 2013.

\bibitem{Batty}
Michael Batty.
\newblock {The size, scale, and shape of cities}.
\newblock {\em Science (New York, N.Y.)}, 319(5864):769--771, 2008.

\bibitem{Bettencourt}
Lu{\'\i}s M A~LM Bettencourt, Jos{\'e}~J Lobo, Dirk~D Helbing, Christian~C
  K{\"u}hnert, and Geoffrey B~GB West.
\newblock {Growth, innovation, scaling, and the pace of life in cities.}
\newblock {\em Proc Natl Acad Sci USA}, 104(17):7301--7306, April 2007.

\bibitem{Pan}
Wei Pan, Gourab Ghoshal, Coco Krumme, Manuel Cebrian, and Alex Pentland.
\newblock {Urban characteristics attributable to density-driven tie formation}.
\newblock {\em Nature Communications}, 4:1961, June 2013.

\bibitem{origins}
Lu{\'\i}s M~A Bettencourt.
\newblock {The origins of scaling in cities.}
\newblock {\em Science (New York, N.Y.)}, 340(6139):1438--1441, June 2013.

\bibitem{Bettencourt2}
Luis Bettencourt and Geoffrey West.
\newblock {A unified theory of urban living}.
\newblock {\em Nature}, 467(7318):912--913, October 2010.

\bibitem{powerlaw}
Michael P~H Stumpf and Mason A~MA Porter.
\newblock {Mathematics. Critical truths about power laws.}
\newblock {\em Science (New York, N.Y.)}, 335(6069):665--666, February 2012.

\bibitem{nee}
S~Nee, N~Colegrave, S~A West, and A~Grafen.
\newblock {The illusion of invariant quantities in life histories}.
\newblock {\em Science (New York, N.Y.)}, 309(5738):1236--1239, 2005.

\bibitem{arbesman}
Samuel Arbesman, Jon~M Kleinberg, and Steven~H Strogatz.
\newblock {Superlinear scaling for innovation in cities}.
\newblock {\em Physical Review E. Statistical, Nonlinear, and Soft Matter
  Physics}, 79(1), January 2009.

\bibitem{Toole:2015bl}
Jameson~L Toole, Carlos Herrera-Yaq{\"u}e, Christian~M Schneider, and Marta~C
  Gonz{\'a}lez.
\newblock {Coupling human mobility and social ties.}
\newblock {\em Journal of the Royal Society, Interface / the Royal Society},
  12(105), April 2015.

\bibitem{georouting}
David Liben-Nowell, Jasmine Novak, Ravi Kumar, Prabhakar Raghavan, and Andrew
  Tomkins.
\newblock {Geographic routing in social networks.}
\newblock {\em Proceedings of the National Academy of Sciences of the United
  States of America}, 102(33):11623--11628, August 2005.

\bibitem{sugden}
Robert Sugden.
\newblock {Rational Choice: A Survey of Contributions from Economics and
  Philosophy}.
\newblock {\em The Economic Journal}, 101(407):751--785, July 1991.

\bibitem{Brockmann:2013bq}
Dirk Brockmann and Dirk Helbing.
\newblock {The hidden geometry of complex, network-driven contagion phenomena.}
\newblock {\em Science (New York, N.Y.)}, 342(6164):1337--1342, December 2013.

\bibitem{Mokhtarian}
Patricia~L Mokhtarian and Cynthia Chen.
\newblock {TTB or not TTB, that is the question: a review and analysis of the
  empirical literature on travel time (and money) budgets}.
\newblock {\em Transportation Research Part A: Policy and Practice},
  38(9-10):643--675, November 2004.

\bibitem{kung}
Kevin~S Kung, Kael Greco, Stanislav Sobolevsky, and Carlo Ratti.
\newblock {Exploring universal patterns in human home-work commuting from
  mobile phone data.}
\newblock {\em PloS one}, 9(6):e96180, 2014.

\bibitem{Simini}
Filippo~F Simini, Marta C~MC Gonz{\'a}lez, Amos~A Maritan, and
  Albert-L{\'a}szl{\'o}~AL Barab{\'a}si.
\newblock {A universal model for mobility and migration patterns.}
\newblock {\em Nature}, 484(7392):96--100, April 2012.

\bibitem{intervening}
S~A Stouffer.
\newblock {JSTOR: American Sociological Review, Vol. 5, No. 6 (Dec., 1940), pp.
  845-867}.
\newblock {\em American Sociological Review}, 5:845--867, 1940.

\bibitem{burchardi}
Konrad~B. Burchardi and Tarek~A. Hassan.
\newblock {The Economic Impact of Social Ties: Evidence from German
  Reunification}.
\newblock {\em The Quarterly Journal of Economics}, 128(3):1219--1271, 2013.

\bibitem{Patel:2014br}
Pragna Patel, Craig~B Borkowf, John~T Brooks, Arielle Lasry, Amy Lansky, and
  Jonathan Mermin.
\newblock {Estimating per-act HIV transmission risk: a systematic review.}
\newblock {\em AIDS (London, England)}, 28(10):1509--1519, June 2014.

\bibitem{Rosenberg:2002ff}
Noah~A Rosenberg and Magnus Nordborg.
\newblock {Genealogical trees, coalescent theory and the analysis of genetic
  polymorphisms.}
\newblock {\em Nature Reviews Genetics}, 3(5):380--390, May 2002.

\bibitem{mattpenrose}
Mathew Penrose.
\newblock {\em {Random Geometric Graphs (Oxford Studies in Probability)}}.
\newblock {Oxford University Press, USA}, July 2003.

\bibitem{Cho:2011io}
Eunjoon Cho, Seth~A Myers, and Jure Leskovec.
\newblock {\em {Friendship and mobility: user movement in location-based social
  networks}}.
\newblock ACM, August 2011.

\bibitem{Wang:2011fz}
Dashun Wang, Dino Pedreschi, Chaoming Song, Fosca Giannotti, and Albert-Laszlo
  Barabasi.
\newblock {\em {Human mobility, social ties, and link prediction}}.
\newblock ACM, August 2011.

\bibitem{gleeson}
James~P Gleeson, Sergey Melnik, Jonathan~A Ward, Mason~A Porter, and Peter~J
  Mucha.
\newblock {Accuracy of mean-field theory for dynamics on real-world networks}.
\newblock {\em Physical Review E}, 85(2):026106, February 2012.

\bibitem{Rozenfeld:2009je}
Hernan~D Rozenfeld, Diego Rybski, Xavier Gabaix, and Hern{\'a}n~A Makse.
\newblock {The Area and Population of Cities: New Insights from a Different
  Perspective on Cities}.
\newblock {\em American Economic Review}, October 2009.

\bibitem{Oliveira:2014gz}
Erneson~A Oliveira, Jose S~Jr Andrade, and Hern{\'a}n~A Makse.
\newblock {Large cities are less green}.
\newblock {\em Scientific Reports}, 4, 2014.

\bibitem{Rozenfeld:2008et}
Hernan~D Rozenfeld, Diego Rybski, Jose S~Jr Andrade, Michael Batty, H~Eugene
  Stanley, and Hern{\'a}n~A Makse.
\newblock {Laws of population growth}.
\newblock {\em Proceedings of the National Academy of Sciences of the United
  States of America}, 105(48):18702--18707, 2008.

\bibitem{Efron}
Bradley Efron and Robert~J Tibshirani.
\newblock {\em {An introduction to the bootstrap}}, volume~57 of {\em
  Monographs on Statistics and Applied Probability}.
\newblock Chapman and Hall, New York, 1993.

\end{thebibliography}

\begin{table*}[ht]
\caption[Travel speed vectors]{Travel speeds of four increasingly developed transport infrastructures. $\mathbf{v}^{(0)}$ represents the trivial case (i.e. no infrastructure). The units are kilometres per hour.}
\begin{tabular*}{\hsize}{@{\extracolsep{\fill}}ccccccccc}
$\mathbf{v}^{(0)}=$&$(4.0,$&$4.0,$&$4.0,$&$4.0,$&$4.0,$&$4.0,$&$4.0,$&$4.0)$  \cr
$\mathbf{v}^{(1)}=$&$(4.0,$&$ 4.8, $&$5.8,$&$ 6.9,$&$ 8.3, $&$10.0, $&$11.9,$&$ 14.3)$  \cr
$\mathbf{v}^{(2)}=$&$(4.0,$&$5.6,$&$7.8,$&$11.0,$&$15.4,$&$21.5,$&$30.1,$&$42.2)$  \cr
$\mathbf{v}^{(3)}=$&$(4.0,$&$6.4 ,$&$10.2,$&$16.4,$&$26.2,$&$41.9,$&$67.1,$&$107.4)$ \cr
\hline 
\end{tabular*}
\label{travelspeeds}
\end{table*} 

\pagebreak

\pagebreak

\clearpage
\appendix
\renewcommand{\theequation}{S\arabic{equation}}
\renewcommand{\thesection}{S\arabic{section}}  
\renewcommand{\thetable}{S\arabic{table}}  
\renewcommand{\thefigure}{S\arabic{figure}}



\section{Derivation of social-tie formulae} 
\subsection{Rank-based tie probability}
Let $\{ Z^{(i)}\}_{i=1}^{N_\text{pop}} $ be a set of positive real-valued random variables representing a single attribute for individuals in a population of size $N_\text{pop}$. We assume that the random variables are independent and identically distributed (i.i.d.) according to some distribution $q(z | z\in\mathbb{R}^+)$. Let $\tau_{ij}$ be the distance matrix specifying the travelling-time distances between individuals, and $S_{ij}$ the temporal social-spheres given by the sets
\begin{equation}
S_{ij} := \{k\, | \, \tau_{ik} < \tau_{ij}\},
\end{equation}
The number of people that are closer to individual $i$ than $j$ as determined from $\tau_{ij}$ is represented by the rank matrix $n_{ij}$, which is simply the cardinality of the temporal spheres, i.e.
\begin{equation}
n_{ij} := |S_{ij}|.
\end{equation}

By design of the proposed social interaction model, in the case where $\tau_\text{max} \rightarrow \infty$, a directed tie from $i$ to $j$ is formed if and only if,
\begin{equation}
z^{(j)} > z^{(i)} \qquad \text{and} \qquad z^{(j)} > \max_{k\in S_{ij}} z^{(k)},\label{conditions}
\end{equation}
where $z^{(i)}$ is a realisation of $Z^{(i)}$ and $S_{ij}$. Let $z_n$ be the maximum value obtained from $m$ random samples from $q(z)$, and let $P_n(\text{c})$ be the probability that $z_n$ satisfies some condition $c$. Then, from \eqref{conditions}, the probability of a directed tie from individual $i$ to $j$ is
\begin{equation}
\begin{split}
\prob(i\rightarrow j) &= \prob(z^{(j)} > z^{(i)}) \times  \int_0^\infty P_1(>z) P_{n_{ij}}( = z)\, \dd z \label{probeq1}\\
&\equiv\int_0^\infty P_1(>z) P_{n_{ij}+1}( = z)\, \dd z.
\end{split}
\end{equation}
Since the population individuals are assumed to be i.i.d. w.r.t. $q(z)$ we have
\begin{subequations}\label{threeeqs}
\begin{align}
P_{1}(>z) &= 1 - q(<z),\\[5pt]
P_{n_{ij}+1}(=z) &\equiv \frac{\dd P_{n_{ij}+1}(< z)}{\dd z}= (n_{ij} +1)\bigl[q(< z)\bigr]^{n_{ij}}.
\end{align}
\end{subequations}
Substituting \eqref{threeeqs} into \eqref{probeq1} and changing variables gives
\begin{equation}
\begin{split}
\prob(i\rightarrow j) &= \int_0^1  \bigl[ q(<z)^{n_{ij}} - q(<z)^{n_{ij}+1}\bigr]\, \dd q(<z)\\
&=\frac{1}{(n_{ij} +2)}.\label{correctrank}
\end{split}
\end{equation}
This is Eq. (8) in the main text. Remarkably, this is independent of the specific attribute $Z$ under consideration. For large rank $n_{ij}$ and up to a constant of proportionality, this expression for the tie-formation probability closely resembles the original rank-based ansatz $\prob(i\rightarrow j) = {1}/{n_{ij}}$ in \cite{Pan, georouting}. The difference here is that the attribute-independence of the tie-probability is an emergent feature rather than a theoretically unsupported assumption of universality. 

\subsection{Social-tie sampling approximation}
Following \cite{Pan} we first assume a uniform population density $\rho$ and travelling-time budget $\tau_\text{max}$. The density is measured in terms of number of individuals per travelling time `volume'. The number of ties $t_i(\rho)$ to a given individual $i$ is
\begin{equation}
\begin{split}
t_i(\rho) &= \int_0^{\tau_\text{max}}\frac{2\pi\tau \rho \,\dd \tau}{\rho\pi\tau^2 + 2} \, \mathrm{d}\tau \\[5pt]
&=  \ln(\rho\pi\tau_\text{max}^2 + 2) - \ln 2\\[5pt]
&= \ln \biggl(\frac{S_i}{2}  + 1\biggr),
\end{split}
\end{equation}
with $S_i= \rho\pi \tau_\text{max}^2$ the size of the temporal social-sphere, i.e. number of nodes reachable from node $i$. Here we are first evaluating the probability of an individual at the origin finding another individual with higher attribute value in a differential spherical volume with the radius is given by the minimum travelling-time distance on the underlying network. Expanding the radius of action on the network, we can then geometrically determine the expected number of ties by integrating up to an attribute-specific limit.
\par
We now drop the dependence on the constant uniform density $\rho$ where the allowance for a heterogeneous distribution is reflected in a varying $S_i$ for different nodes $i$. We replace $t_i(\rho)$ with $t_i$ to indicate this transition. The total number of ties $T$ in the population is, then, simply
\begin{equation}
\begin{split}
T = \sum_{i=1}^{N_\text{pop}} t_i &= N_\text{pop}\Biggl[\frac{1}{N_\text{pop}} \sum_{i=1}^{N_\text{pop}} \ln\biggl( \frac{S_i }{2}+1\biggr)\Biggr] \\
&\leq N_\text{pop}\ln \Biggl[ \frac{1}{N_\text{pop}} \sum_{i=1}^{N_\text{pop}}  \biggl( \frac{S_i }{2}+1\biggr)\Biggr] \\
&= N_\text{pop}\ln \biggl( \frac{\bar{S}}{2} + 1\biggr),\label{fulleqq}
\end{split}
\end{equation}
where $\bar{S} = 1/{N_\text{pop}}\sum_{i=1}^{N_\text{pop}} S_i$ is the population average of the number of reachable nodes. The inequality is due to Jensen's inequality and the concavity of the logarithmic function.

However obtaining the full set $\{t_i\}_{i=1}^{N_\text{pop}}$ is neither possible or practical for typically-sized cities. We therefore take a sample of $N_s$ points. Defining $\alpha \equiv N_\text{pop}/{N_s}$, if the sample is representative of the population, we have
\begin{equation}
\frac{\bar{S}}{ \bar{n} } = \alpha,
\end{equation}
where $\bar{n}$ is the average of the number of reachable nodes within the sample set, i.e. $\bar{n} = (1/N_s)\sum_{i=1}^{N_s}n_i$. We have
\begin{equation}\label{sampleref}
\begin{split}
N_\text{pop}\ln \biggl(\frac{\bar{S}}{2} + 1\biggr) & = N_\text{pop}\ln \biggl(\frac{\alpha\bar{n}}{2} + 1\biggr)\\
&=N_\text{pop}\ln\Biggl[\frac{\alpha\bar{n}}{2}\biggl(1+\frac{2}{\alpha\bar{n}}\biggr)\Biggr]\\
&= N_\text{pop}\Biggl[\ln \frac{\alpha}{2}  + \ln \bar{n} + \ln\biggl(1 + \frac{2}{\alpha\bar{n}}\biggr)\Biggr]\\
&\approx  N_\text{pop}\biggl(\ln \frac{\alpha}{2} + \ln \bar{n} + \frac{2}{\alpha\bar{n}}\biggr)\\
& \geq N_\text{pop}\Biggl[\ln \frac{\alpha}{2}  + \frac{1}{N_s}\sum_{i=1}^{N_s}\ln n_i+ \biggl(\frac{2}{\alpha\bar{n}}\biggr)\Biggr].
\end{split}
\end{equation}
Combining \eqref{fulleqq} and \eqref{sampleref} we expect the two inequalities to cancel out approximately, giving
\begin{equation}\label{sampformula}
T \approx N_\text{pop} \Biggl[ \ln \biggl(\frac{N_\text{pop}}{2N_s}\biggr) + \frac{1}{N_s}\sum_{i=1}^{N_s} (\ln n_i )\Biggr] + \frac{2N_s}{\bar{n}}.
\end{equation}
This is Eq. (10) in the main text.

\subsection{Local connectivity}
The local connectivity is defined as half the sum of incoming and outgoing ties from a given location. Let $T_i$ represent the local connectivity of the location of individual $i$. By definition, we have
\begin{equation}\label{tofrom}
T_i = \frac{1}{2}\bigl(T_i^\text{from} + T_i^\text{to}\bigr), \qquad \text{with} \qquad \sum_{i=1}^{N_\text{pop}} T_i= T.
\end{equation}
As in the case of global connectivity, the key is to approximate $T_{i}$ without having access to the population distance matrix. We estimate the outgoing and incoming contribution separately, beginning with the outgoing component $T_i^\text{from}$.

Following the reasoning behind \eqref{sampleref}, we have
\begin{equation}\label{from}
T_i^\text{from} = \frac{1}{2}\ln \left(\frac{\alpha n_i}{2} + 1\right).
\end{equation}

Quantifying the incoming ties is less straightforward as there is no simple scaling from the sample. Instead we perform the approximation in three stages on the basis of several reasonable assumptions. First, for two individuals $i$ and $j$ in our population sample, we approximate the true population rank $n_{ij}$ from the sample rank $\hat{n}_{ij}$, i.e.
\begin{equation}\label{estrank}
n_{ji} = (\alpha-1) + \frac{1}{2}(\alpha - 1) + \alpha \hat{n}_{ji} = \alpha\left(n_{ji} + \frac{3}{2}\right) -\frac{3}{2},
\end{equation} 
where the three terms in the sum are, respectively, the scaled contributions from individuals $i$, $j$ and the $\hat{n}_{ji}$ intervening samples\footnote{NB: $n_{ij}\neq n_{ji}$. We assume throughout that individual $i$ is the seeker, i.e. the \textit{recipient} of incoming ties.}. Second, from \eqref{correctrank} and \eqref{estrank}, the probability of an directed tie from $j$ to $i$ is 
\begin{equation}\label{probest}
\prob(j\rightarrow i) = \frac{1}{\alpha(n_{ji} + \frac{3}{2}) + \frac{1}{2}}.
\end{equation}
A first approximation of the expected total incoming ties is the appropriately-scaled sum of all incoming tie probabilities from our sample, i.e.
\begin{equation}
T_i^\text{to} = \alpha\sum_{\substack{j=1\\j\neq i}}^{N_s}\prob(j\rightarrow i).
\end{equation}
However, imposing the consistency criteria $\sum_{i=1}^{N_s} T_i^\text{from} = \sum_{i=1}^{N_s} T_i^\text{to}$ requires a third step of scaling \eqref{probest} appropriately. Therefore, we have
\begin{equation}\label{to}
T_i^\text{to} = {\gamma\alpha}\sum_{\substack{j=1\\j\neq i}}^{N_s}\prob(j\rightarrow i),
\end{equation}
with 
\begin{equation}
\gamma = \frac{\sum_{i=1}^{N_s} T_i^\text{from}}{\alpha\sum_{i=1}^{N_s}\sum_{{j=1, j\neq i}}^{N_s}\prob(j\rightarrow i)}.
\end{equation}
Substtituing \eqref{from} and \eqref{to} into \eqref{tofrom}, we obtain Eq. (11) in the main text.

\section{Induced network structure from social dynamics}
In this paper we have constructed a probability model for generating social-tie networks where the edges denote deliberate (i.e. planned as opposed to random encounters) face-to-face interactions.  It is worth reemphasising that the networks throughout are themselves unobserved structures, which compels one to average over all possible networks. In this section we provide three instances of how our model can be coaxed to provide additional secondary expected network summary statistics. This is in addition to the expected number of interactions, i.e. the expected number of network edges, which we examined in the section above, and have showed in the main text to be a sufficiently strong predictor for several urban indicators. Specifically, we look at network heterogeneity, multilevel network structures, and spatial extent of spatial networks.

\subsection{Network heterogeneity}
\begin{figure*}[ht] 
\begin{center}
\centerline{\includegraphics[width=.9\textwidth]{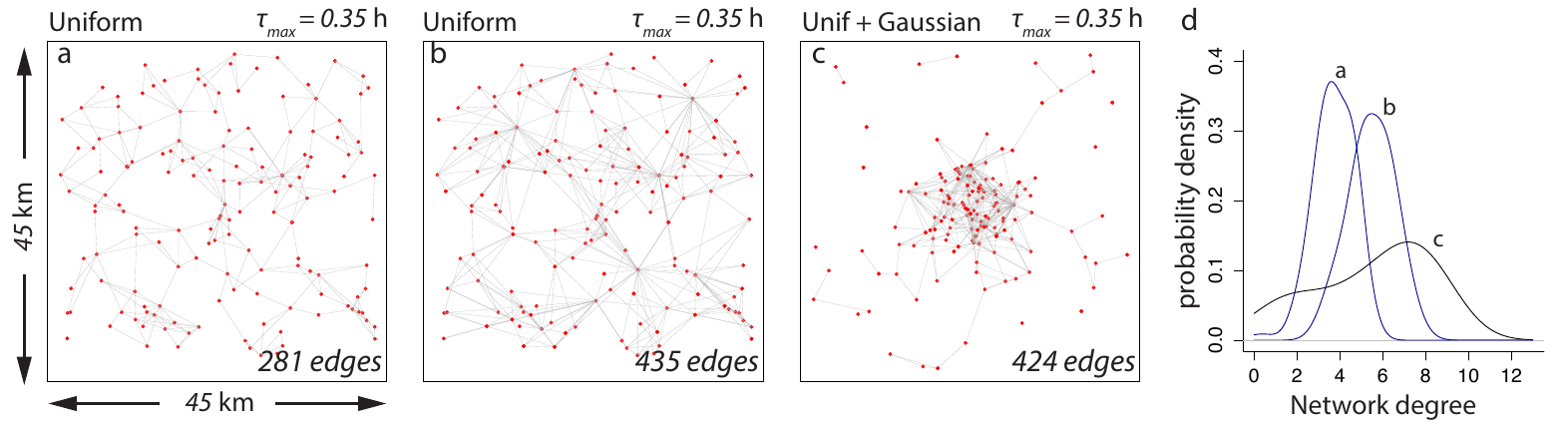}}
\caption[Network heterogeneity]{\textbf{Emergence of network structure}. \textbf{a,b,c,} Simulated city interaction network examples. The red nodes represent individuals while the network edges indicate a directed social-tie (direction not specified). The nodes in networks \textbf{a} and \textbf{b} are uniformly distributed while those in network \textbf{c} is sampled from a $(1/3, 2/3)$-weighted mixture of a uniform and Gaussian distribution with component standard deviation of 4 km.  \textbf{d,} The network degree distributions, averaged over 120 different (and random) attribute-value distributions, for the three simulated cities.}\label{networkhet}
\end{center}
\end{figure*}

We focus on the impact that different population distributions and travelling-time budgets have on the expected network degree distributions. We simulate three cities following the procedure outlined in the section above. The first two cities are networks with 150 nodes uniformly distributed with average $\tau_\text{max}=0.35$ and $\tau_\text{max}=0.35$ respectively, while the third is a network with 150 nodes sampled from a $(1/3, 2/3)$-weighted mixture of a uniform and Gaussian distribution with component standard deviation of 4 km and average $\tau_\text{max}=0.5$. The travelling-time budgets were chosen such that the second and third networks possess a similar number of edges. In both cities the transport infrastructure is assumed to have three modes and is represented by the speed vector 
\begin{equation} \label{speedvectorthree}
v' = (4.0, 15.0, 15.0, 33.0, 33.0, 33.0, 33.0, 33.0).
\end{equation}
As before, the values have units of kilometres per hour and here the three values represent the average speeds of walking, bus, and metro travel. Three example networks for a given population distributions of attribute values are shown in Figure \ref{networkhet}. First we observe, somewhat trivially, that for a single city an increase in $\tau_\text{max}$ can lead to an increase in number of edges. Second, for similar spatial distributions, we see that the network degree distributions of two cities can be markedly different, even in the case where the number of edges are similar. Here we compare the expected degree distribution, taking the average over 120 random attribute-value ranking distributions. As shown in Figure \ref{networkhet}, the city with a dense centre has a significantly higher level of network heterogeneity than the uniformly distributed city. 

\subsection{Spatial extent of spatial networks}
Since the underlying interaction networks behind the connectivity measure are spatial networks, it can be useful to examine the impact of urban infrastructure changes not just on overall and local connectivities, but on the spatial nature of those changes. In this section we use the example of London Crossrail from the main text. There we calculate both the impact on total connectivity and its local spatial variations. Here, we go one step further by predicting the expected distribution of the interaction network edge lengths in the city of London (in terms of Euclidean spatial distances) before and after the construction of London Crossrail. 
\par
\begin{figure*}[ht] 
\begin{center}
\centerline{\includegraphics[width=.7\textwidth]{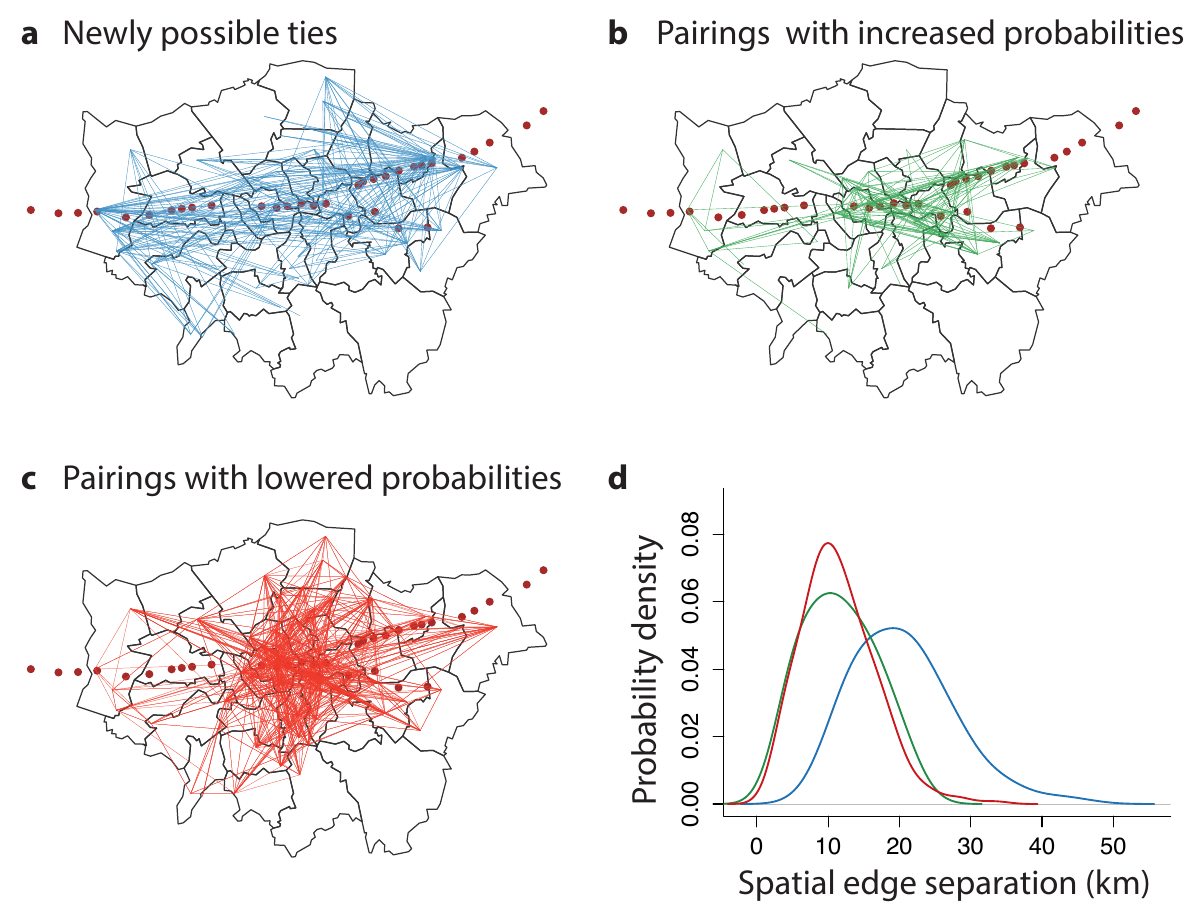}}
\caption[Network decomposition]{\textbf{Crossrail effect on London interaction network}. \textbf{a}, Newly possible interaction edges, \textbf{b}, existing possible interactions that have increased in probability, \textbf{c}, existing  possible interactions that have decreased in probability. The three networks are taken from a subnetwork with 70 nodes. \textbf{d}, Expected distribution of the interaction network edge lengths for the three classes of interactions. The edge lengths are given in terms of the spatial Euclidean distances between nodes.}
\label{networkLondon}
\end{center}
\end{figure*}

The results are presented in Figure \ref{networkLondon}. We make three observations. First, the newly possible interactions (i.e. those with probability zero in the absence of Crossrail) tend to have higher average edge lengths than existing interactions. Second, the existing connections that are upgraded or downgraded in probability seem to have identical spatial length distribution. Third, the increases tend to occur along the new railway route while the decreasing edges tend to have one or both nodes in the orthogonal dimension (i.e. north-south corridor). The conclusion here is highly intuitive: apart from new connection possibilities between regions that are otherwise separated by large spatial, the changes in interaction probabilities depends less of distance between nodes than the nodes' locations relative to the new infrastructure.

\section{Details of empirical validation examples}
102 US Metropolitan Statistical Areas (MSAs) were chosen on the basis of the availability of HIV infection rate, GDP, and spatial population distribution data. 

\subsection{Data sources} The population statistics and density profiles of the 102 US MSAs are obtained from the U.S. Census Bureau\footnote{\textit{2010 Census of Population and Housing}, \textit{2010 U.S. Metropolitan Statistical Area Distance Profiles}, www.census.gov}. HIV infection and prevalence data are obtained from the United States Centers for Disease Control and Prevention\footnote{US Centers for Disease Control and Prevention. \textit{HIV Surveillence Report}, 2011; vol.23. www.cdc.gov/hiv/topics/surveillance/resources/reports/. Feb 2013.}. Travel times between city locations are obtained using Microsoft BING maps\footnote{www.microsoft.com/maps/} and for car journeys originating at 1200h local-time on 13th Dec 2013.

Given the marginal radial distribution of the population, we assume a circular symmetry about the central city hall location and sample a set of 1000 points for each US MSA. One can drop this assumption and obtain more accurate and precise population distribution data, for instance from detailed local census and other open data sources. The total number of ties $T_i$ in each MSA $i$ is then calculated  by applying \eqref{sampformula} to the travelling-time distance matrices obtained from online mapping resources. The mode of transport here is restricted to travel by roads -- an assumption that is reasonable for many US MSAs.

In the example of HIV infection rates, the relevant number of ties are encounters, $T'$, between HIV-positive and HIV-negative individuals, rather than the total number of ties. We therefore scale the total no. of ties by the mixed-ties proportions, giving
\begin{equation}
T' = \frac{2H(N-H)}{N(N-1)}T,
\end{equation}
where $N$ is the population size of a given MSA and $H$ the number of individuals diagnosed as HIV-positive, which we take to be equal to the reported number of HIV-positive individuals in the population.

\subsection{Robustness of $\tau_\text{max}$ estimation}
In this section we gauge the robustness of the maximum likelihood travelling-time budget estimates $\tau^{mle}_\text{max}$ obtained for the HIV-infection rates and GDP-related attributes by constructing confidence intervals (C.I.) around the respective point estimates. We present two versions: a bootstrap C.I. and a C.I. based on the asymptotic variance of the maximum likelihood estimator in terms of the observed Fisher information.

Using $N_{B}=1000$ bootstrap replicates of the original set of 102 US cities, we repeat the linear fits of the urban indicators to tie-density. We obtain a set of bootstrap travelling-time maximum likelihood estimates $\{{\tau}^{mle}_{\text{max},i}\}_{i=1}^{N_B}$ which then provides a bootstrap confidence interval $C_\text{boot}$ in terms of the empirical quantiles \cite{Efron}. 

Next, we assume that the residues of the $\log\,U-\log\,T$ linear fit $\log \, U = g_{\tau_\text{max}}(\log \,T)$ are normally distributed with $\hat{s}$ be the sample standard deviation of the maximum likelihood fit $g_{\tau^{mle}_\text{max}}$. We further assume that the data points are independent, whereby the likelihood is
\begin{equation}
\mathcal{L}(\tau_\text{max}) = \prod_{i=1}^{n} f_{\tau_\text{max}}(\log\,U_i).
\end{equation}
$f_{\tau_\text{max}}(\log \, U_i)$ the univariate normal density function with mean $g_{\tau_\text{max}}(\log \, T_i)$ and variance $(\frac{\hat{s}}{\hat{s}-1})^2$.
 The observed Fisher information $I$ is then
\begin{equation}\label{FI}
I = {-\frac{d^2 L(\tau_\text{max})}{d^2 \tau_\text{max}}} \,\bigg\vert_{\tau_\text{max}=\tau^{mle}_\text{max}},
\end{equation}
where $L = \log\,\mathcal{L}$ is the log-likelihood. Practically, we obtain \eqref{FI} through a series of Gaussian Process fits though the set of emprical data points $\{\tau_\text{max}, L(\tau_\text{max})\}$. In the asymptotic limit, the maximum likelihood estimator is normally distributed with variance $-1/I$ which is used to define the C.I. $C_\text{mle}$. Strictly speaking, the asymptotic distribution is clearly not normal as the parameter $\tau_\text{max} >0$. However, at least for the GDP attribute, the maximum likelihood estimate is sufficiently away from the zero boundary for this to be a reasonable assumption.

\begin{figure*}
\begin{center}
\centerline{\includegraphics[width=.7\textwidth]{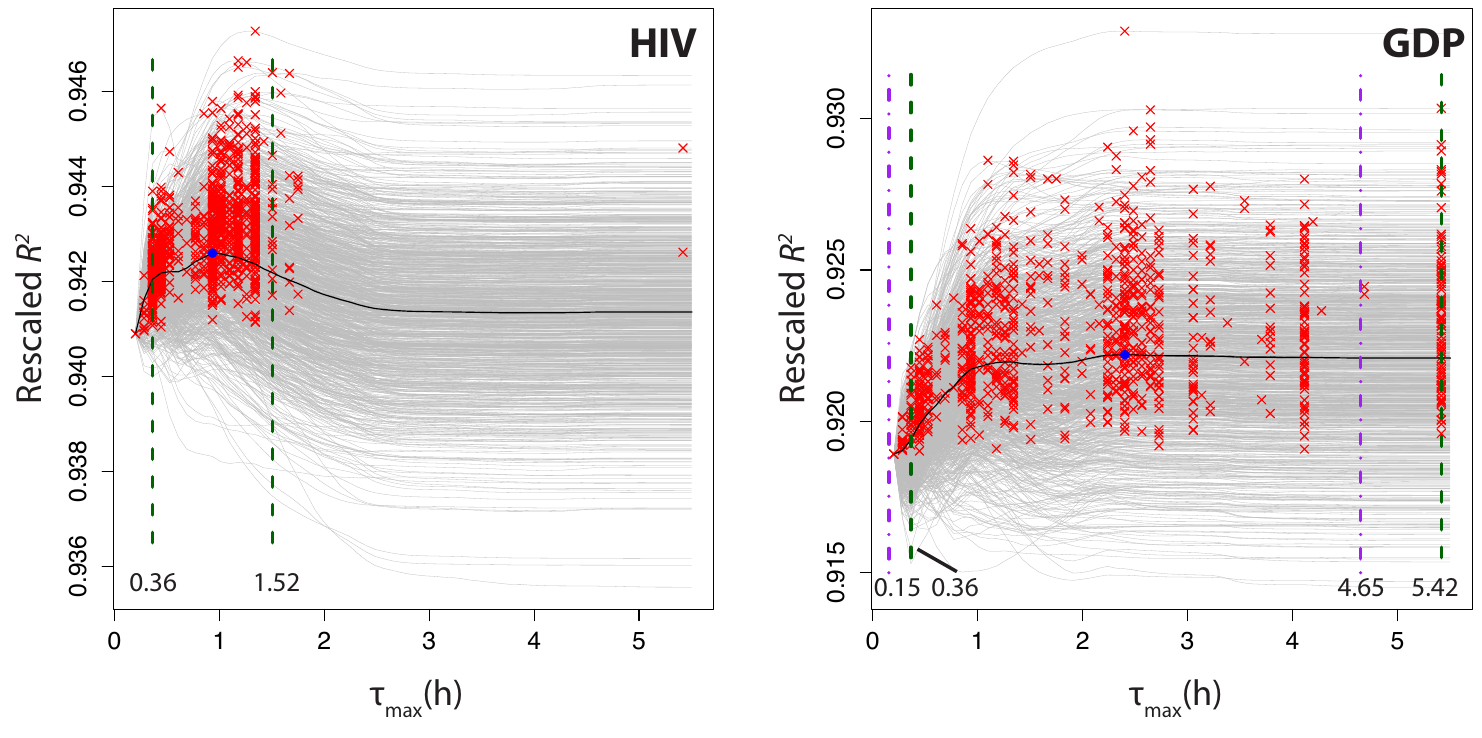}}
\caption[Robustness of $\tau_\text{max} estimates]{\textbf{Robustness of $\tau_\text{max}$ estimates}. Plots of rescaled-$R^2$ values of $\log\, U-\log\,T$ linear fits as a function of $\tau_\text{max}$. The black solid line is the curve using the original dataset of 102 cities. Each of the 1000 grey curves represents the $R^2$ values of a separate bootstrap sample of the original data, rescaled such that $\tau_\text{max}$ matches the value from the original dataset. The red crosses indicate the maximum $R^2$ values of the bootstrap curves and the blue circle the same for the original curve. The green dashed and purple dot-dashed lines indicate the 95\% bootstrap confidence interval and the observed Fisher information-derived confidence interval respectively. }
\label{robustnessfig}
\end{center}
\end{figure*}

The C.I.s are illustrated in Figure \ref{robustnessfig} for both attributes. From the analysis we have the 95\% C.I. $C_\text{boot}=[0.36, 1.52]$ and $C_\text{boot} =[0.36,5.42]$ ($C_\text{mle}=[0.15, 4.65]$) for the HIV infection rates and GDP-related attributes respectively. While the $\tau_\text{max}$ estimate for the HIV infection rates attribute is fairly robust, the C.I. for the GDP estimate spans $>4$ hours. This behaviour confirms the intuition that the GDP indicator pertains, in reality, to an amalgamation of many attributes with varying sizes of $\tau_\text{max}$. For instance, a typical city inhabitant is unlikely to patronise a laundromat more than a few minutes from home; on the other hand, the same person is probably willing to endure a long commute across the city for a one-off visit to a unique theme park, say. Both activities contribute to GDP, and this difference is reflected in the wide span for the $\tau_\text{max}$ estimates.

\section{Details of HS2 and London Crossrail analysis}\label{HS2sec}
\subsection{Data sources} London and Birmingham demographic profiles and geographical details are obtained from the Greater London Authority\footnote{http://data.london.gov.uk/datastore} and the Birmingham City Council\footnote{www.birmingham.gov.uk} respectively. Details of HS2, including routes, station locations and travel speeds are obtained from the High Speed Two Limited\footnote{www.hs2.org.uk}. London Crossrail station and travel times are obtained from Transport for London\footnote{www.crossrail.co.uk}. Current travelling times between city locations are obtained using Microsoft BING maps\footnote{www.microsoft.com/maps/}.

We obtain geographic samples from the cities of London and Birmingham, UK from two-dimensional (weighted) kernel density estimates (KDE) of the population spatial distributions. The central locations of the 32 boroughs in London and 40 wards in Birmingham are treated as data points with weights proportional to the local population sizes. We use a Gaussian kernel with bandwidth equal to 1.2 times the radius of a circle with area equal to the local borough or ward for each data point. The population of each city is then sampled from this weighted mixture of Gaussians. We have a total of 1000 and 128 location samples for London and Birmingham respectively.

The travelling time distance matrices used represent, for the majority of point-pairs, public transport travelling time. In the absence of public transport data between two locations, we assume that the relevant journey is taken by car. As for the US MSA examples, the data is obtained from online mapping resources. We selected a departure time of 1200 on 12th December 2013.

For the HS2 example, we assume a single interchange station in each city (London Euston, and Curzon Street station in Birmingham). The travelling time between locations in each city is a sum of the travelling times to each station and the published journey time between the two cities (we do not factor in waiting times, delays, etc.). 

There are 36 stations on the London Crossrail network, with an additional 12 in the CrossRail 2 (metro-only option) extension. The improved travelling time between two London locations is the sum of the travelling times from the origin and destinations to their respective closest (by time) Crossrail stations and the published station-to-station journey time.


\begin{figure*}[ht]
\begin{center}
\centerline{\includegraphics[width=.7\textwidth]{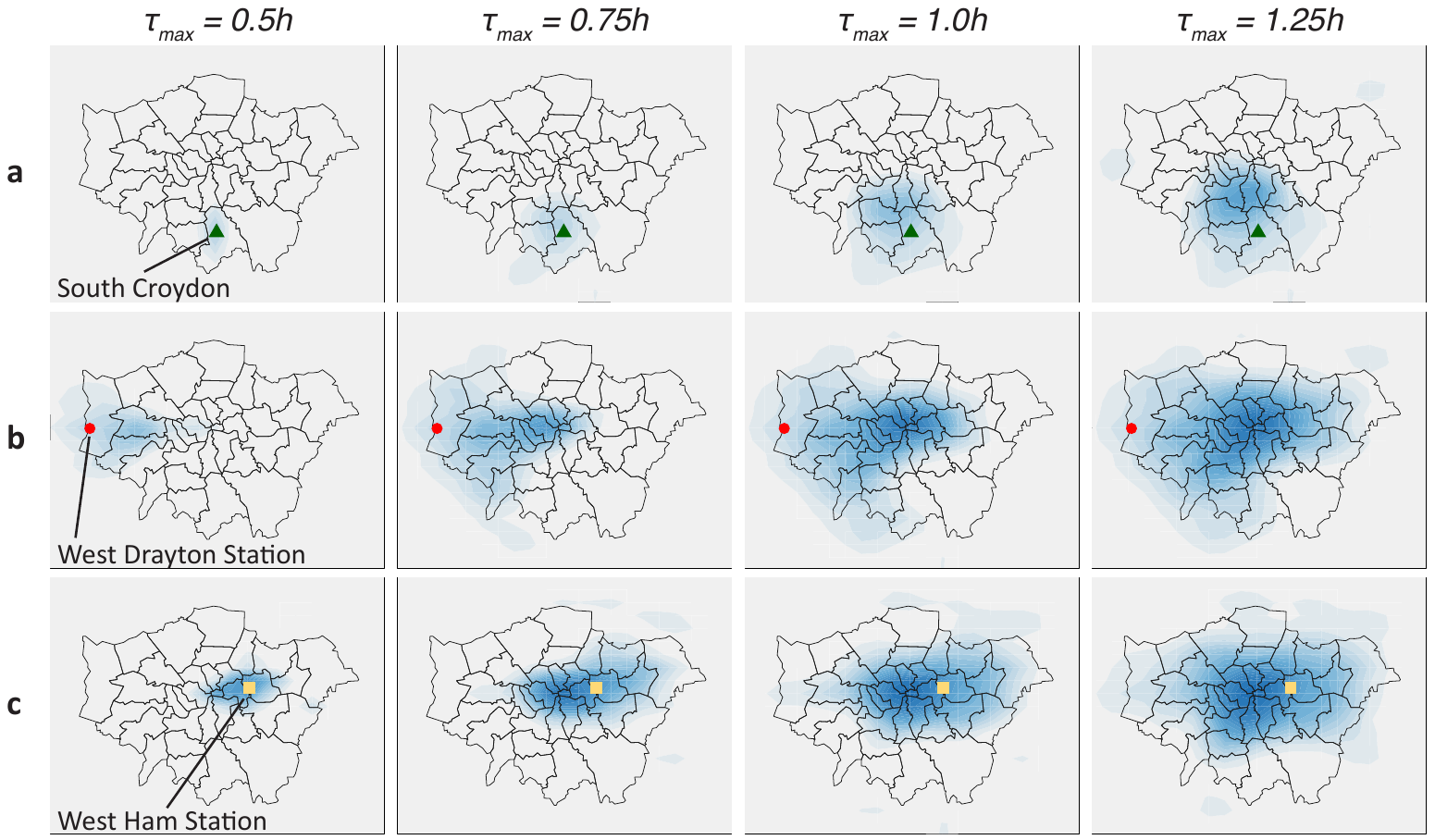}}
\caption[Spheres of influence]{\textbf{Density maps of \textit{travelling-time social spheres} in London (with Crossrail) as a function of $\tau_\text{max}$ and location.} The coloured square, circle and triangle represent example central, western, and southern locations in London respectively. The contour maps represent kernel density estimates of samples ($N_s = 1000$) within the indicated travelling-time distance budget. The western location in \textbf{b} lies directly on a Crossrail station (West Drayton station), while the southern location (South Croydon station) in \textbf{c} is chosen to illustrate a relatively inaccessible location in the city. See \textit{SI} Section \ref{HS2sec} for details of the construction of the distance matrices used.}\label{KDE}
\end{center}
\end{figure*}



\end{document}